\documentclass[aps,prl,nofootinbib,showpacs,twocolumn]{revtex4-1}%
\usepackage{amssymb,latexsym,mathrsfs}
\usepackage{amsfonts}
\usepackage{mathrsfs}
\usepackage{amsmath}
\usepackage{graphicx}
\usepackage{srcltx}
\usepackage{color}
\usepackage{cancel}
\begin{document}

\title{Fixed-Energy Sandpiles Belong Generically to Directed Percolation}
\author{Mahashweta Basu$^1$, Urna Basu$^1$, Sourish Bondyopadhyay$^1$, P. K. Mohanty$^1$, and Haye Hinrichsen$^2$}
\affiliation{$^1$TCMP Division, Saha Institute of Nuclear Physics,
1/AF Bidhan Nagar, Kolkata 700064, India.\\
$^2$Universit\"at W\"urzburg, Fakult\"at f\"ur  Physik  und Astronomie, 97074  W\"urzburg, Germany.}

\begin{abstract}
Fixed-energy sandpiles with stochastic update rules are known to exhibit a nonequilibrium phase transition from an active phase into infinitely many absorbing states. Examples include the conserved Manna model, the conserved lattice gas, and the conserved threshold transfer process. It is believed that the transitions in these models belong to an autonomous universality class of nonequilibrium phase transitions, the so-called Manna class. Contrarily, the present numerical study of selected   (1+1)-dimensional models in this class suggests that their critical behavior converges to directed percolation after very long time, questioning the existence of an independent Manna class.
\end{abstract}

\maketitle

Self-organized criticality (SOC) was introduced in the late 80's as an attempt to explain the ubiquitous variety of scale-invariant phenomena in nature~\cite{SOCReview}. Paradigmatic examples are sandpile models (see, e.g., \cite{Chessa}), where sand is accumulated on a  slow time scale and relaxed in form of sudden avalanches on a fast time scale. The interplay of slow driving and fast relaxation combined with dissipation at the boundaries drives such systems towards a scale-invariant state without any fine tuning of parameters. As a hallmark of SOC, one observes power-law distributed avalanche sizes.

Later it became clear~\cite{Tang,Dickman} that SOC is closely related to nonequilibrium phase transitions into infinitely many absorbing states~\cite{Henkel}. To make this relation more explicit fixed energy sandpiles (FES) were introduced, where grains on a periodic lattice follow the same local toppling rules as the original sandpile models without any drive or dissipation. One of the best studied models is the fixed energy version of the nondeterministic Manna sandpile model~\cite{Manna,Alava}, the discrete conserved Manna model (DCMM). The transition in this model was found to be different from directed percolation (DP) and other previously known universality classes \cite{Munoz}. This point of view was bolstered by the discovery that various other models with conservation laws such as the conserved lattice gas model (CLG), the conserved threshold transfer process (CTTP)~\cite{Lubeck,Rossi}, and the Maslov-Zhang sandpile~\cite{Maslov}, exhibit the same type of universal critical behavior, constituting the so-called Manna class (MC) of absorbing phase transitions with a conserved field which is  considered today as firmly established~\cite{Henkel}. 

The paradigm of an independent Manna class, however, has always been overshadowed by several disturbing observations~\cite{SM}. The reported estimates for the critical exponents of the MC are quite scattered and show anomalous scaling behavior. For example, the decay of activity from a homogeneously active state, $\rho_a(t) \sim t ^{-\alpha}$, and of the survival probability of an avalanche starting with a single seed $P_s(t) \sim t^{-\delta}$ are characterized by different critical exponents $\alpha \ne \delta.$ On the other hand one finds that the order parameter exponent $\beta$ and the survival probability exponent $\beta'$ are equal within error bars \cite{Heger2}. Moreover, the upper critical dimension $d_c=4$ and the mean-field exponents 
for the MC~\cite{Heger2} are same as in DP. In addition, the corresponding SOC models are known to be unstable against specific perturbations and generically flow to DP \cite{Mohanty}. 

In this Letter, we suggest that an independent Manna class does not exist. Instead, we find that the apparent nonDP behavior, which was seen in many numerical studies, is a transient phenomenon, meaning that all models of this type are expected to show a  DP  critical behavior after very long time. 

%========================================================
\begin{figure}[t]
\centering\includegraphics[width=85mm]{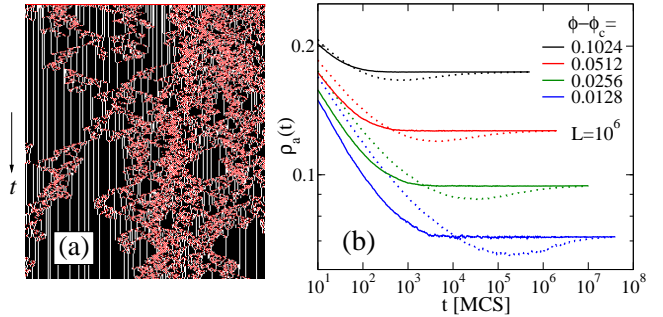}
\vspace{-4mm}
\caption{ DCMM: (a) Typical time evolution, where single and active sites are marked by black and red pixels, respectively. (b) Decay in the supercritical phase for random initial conditions (dotted lines) and natural initial states (solid lines). }
\label{fig:fig1}\vspace*{-.5 cm}
\end{figure}
%========================================================

\paragraph{Discrete Model:}  
To support this hypothesis, we first revisit the DCMM, showing that its critical behavior depends crucially on correlations in the initial state. We restrict our study to (1+1) dimensions, where possible discrepancies between the MC and DP are expected to be most pronounced. The DCMM is defined on a chain with $L$ sites and periodic boundary conditions, where each site $j = 1, 2, \dots , L$ is occupied by $n_j$  particles. Sites with $n_j\ge  2$ particles are declared as active,  labeled by a flag $s_j=1$, while empty and singly occupied sites are inactive ($s_j=0$). The model evolves by parallel updates, redistributing particles at active sites independently among randomly selected nearest-neighbor sites so that the total number of particles $N=\sum_j n_j$ is conserved [see Fig.~\ref{fig:fig1}(a)]. The DCMM exhibits a continuous transition from an active phase into infinitely many absorbing states controlled by the average density of particles $\phi = N/L$ in the initial state, where the density of active sites  $\rho_a=\langle s_i\rangle$ plays the role of an order parameter. 

\paragraph{Random initial conditions:} 
Studying homogeneously active initial states with a given density $\phi$, most authors  used to distribute $N$ particles randomly on the chain. In the active phase $\phi>\phi_c$ one expects the density of active sites $\rho_a(t)$ to cross over monotonically from an algebraic decay to a constant value. Surprisingly, we find that $\rho_a(t)$ first reaches a minimum, then increases and finally saturates at a stationary value [see Fig. \ref{fig:fig1}(b)]. Such a nonmonotonic undershooting is quite unusual for absorbing phase transitions. Moreover, the curves cannot be collapsed onto a single one, indicating the presence of several time scales~\cite{SM}. We believe that this circumstance is the origin of various numerical inconsistencies reported in the literature. 

%========================================================
\begin{figure}[t]
\centering\includegraphics[width=87mm]{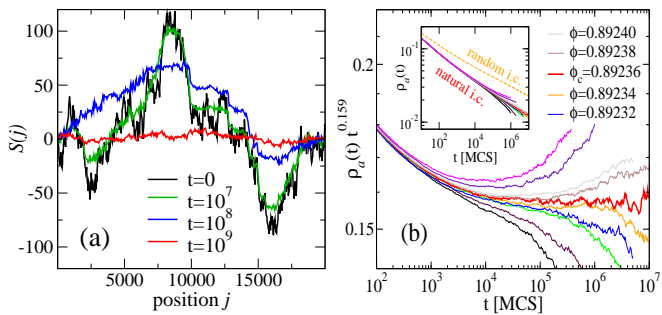}
\vspace{-6mm}
\caption{ DCMM: (a) Cumulative background density for random initial conditions captured at different simulation times. (b) Decay of the density of active sites for natural initial conditions on a system with $L=2^{18}$ sites.}
\label{fig:fig2}
\end{figure}
%========================================================

\paragraph{Explaining the undershooting:}
As demonstrated in Fig.~\ref{fig:fig1}(a), the dynamics of active sites takes place on a heterogeneous background of immobile particles. For random initial conditions (i.c.) this background is highly disordered so that the spreading process is expected to behave like DP with spatially quenched disorder~\cite{disorder}, slowing down the decay of $\rho_a(t)$ and lowering its  value in the active phase. However, the disorder of the background is not quenched but 
 gets slowly modified by the process itself. We find that this feedback gradually homogenizes the background on a very slow time scale, leading to a subsequent increase of $\rho_a(t)$. The gradual removal of disorder is visualized in  Fig. \ref{fig:fig2}(a), where we plotted the cumulative sum $S(j)=(\sum_{i=1}^j n_i) - Nj/L$ which measures the excess of particles to the left of site~$j$ compared to the expected average. As can be seen, the pronounced density fluctuations of the random initial state (black curve) are gradually leveled out, producing an almost flat profile (red line) after very long time.

\paragraph{Natural homogeneous initial states:}
In presence of undershooting, the additional curvature in  $\rho_a(t)$ can easily lead to an erroneous estimate of $\phi_c$ and the critical exponents~\cite{SM}. As one of our main results, we show that the observed undershooting in the supercritical phase can be avoided by preparing natural initial states. Following an idea introduced earlier in the context of seed simulations~\cite{Natural}, we first let the process run until it becomes stationary after the undershooting. As diffusion is known to be the conjugate field in the DCMM~\cite{Lubeck2}, we  then  reactivate the system by allowing all the particles to diffuse for a single Monte Carlo sweep (MCS), which restores a high  homogeneous activity without destroying the natural long-range correlations 
in the background. After reactivation we measure $\rho_a(t)$ as usual. As shown in Fig. \ref{fig:fig1}(b), this resolves the problem of undershooting. 

Figure ~\ref{fig:fig2}b shows that the temporal decay of $\rho_a(t)$ for random and natural i.c. is in fact very different. For the latter we find the critical point $\phi_c=0.89236(3)$, which differs significantly from the previously reported estimate $\phi_c = 0.89199(5)$ obtained with random~i.c.~\cite{Lubeck,SM}. For the decay exponent, which was previously estimated by $\alpha = 0.141(24)$, we get a much larger value $\alpha=0.159(3)$ for natural i.c. which is in agreement with DP.

%========================================================
\begin{figure}[t]
\centering\includegraphics[width=87mm]{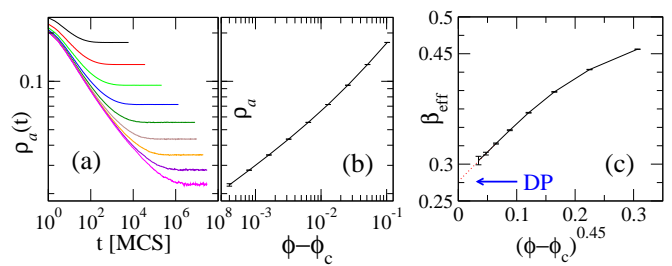}
\vspace{-7mm}
\caption{ DCMM: (a) Decay of the $\rho_a(t)$ with natural initial conditions and $L=2^{18}$ sites for $\phi$-$\phi_c$=0.0004, 0.0008, \dots 0.1024. Here $\phi_c=0.89236$ obtained from Fig. \ref{fig:fig2}(b). (b) Corresponding stationary densities. (c) Extrapolation of the static exponent $\beta$.}
\label{fig:fig3}
\end{figure}
%========================================================

\begin{figure}[t]
\centering\includegraphics[width=87mm]{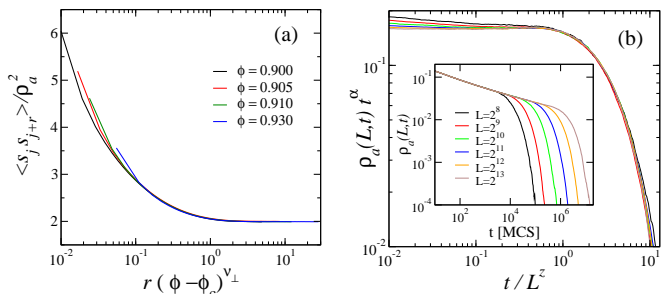}
\vspace{-8mm}
\caption{ DCMM : (a) Spatial correlation function in the stationary state  with system size $L=2^{11}$. (b) Finite-size scaling at criticality with natural homogeneous i.c. for $L =2^8$ to $2^{13}$ and corresponding data collapse. }
\label{fig:fig4}
\end{figure}

\paragraph{Stationary state:}
Figure ~\ref{fig:fig3}a shows the saturation of $\rho_a(t)$ in the supercritical phase using natural initial conditions. As can be seen in panel (b), the stationary density plotted against $\Delta=\phi-\phi_c$ displays a slight curvature. Approaching criticality the effective exponent $\beta_{\rm eff}$ (the slope between adjacent data points) exhibits a drift from which we can safely conclude that $\beta<0.31$. Plotting $\beta_{\rm eff}$ against  $\Delta^b$ with a heuristically determined exponent  $b=0.45$ in panel (c), the data points seem to follow an asymptotically straight line, allowing us to extrapolate $\beta_{\rm eff}$ visually to criticality. As indicated by the blue arrow, the result is again compatible with DP. Likewise, by plotting $\rho_a(t)/\rho_a$ against $t\Delta^{\nu_\parallel}$ and searching for a data collapse (not shown here) we estimate $\nu_\parallel=1.75(5)$. In addition, we estimated the exponent $\nu_\perp$ by measuring the correlation function $C(r,\Delta) = \langle s_j s_{j+r}\rangle$ in the stationary state, which obeys the scaling form $C(r,\Delta) =\rho_a^2 {\cal G}( r\Delta^{\nu_\perp})$ near criticality. A data collapse in Fig.~\ref{fig:fig4}(a) gives the estimate $\nu_\perp= 1.095(5).$ Both exponents $\nu_\parallel$ and $\nu_\perp$ are in good agreement with DP. 

%========================================================
%========================================================

\paragraph{Finite-size scaling:}
The dynamical exponent $z=\nu_\parallel/\nu_\perp$ can be determined by finite-size simulations at criticality, where the density of active sites is expected to obey the scaling form $\rho_a(t,L)= t^{-\alpha}  {\cal F}\left(\frac {t}{L^z}\right)$. Again a conventional data collapse in Fig. \ref{fig:fig4}(b) leads to the estimate $ z=1.51(5)$ which is somewhat smaller than the corresponding DP value $1.58$ but larger than previously reported estimates.

%========================================================
\begin{table}[b] \begin{center} \begin{small}
\begin{tabular}{|l|c|c|c|c|c|}\hline
Model &$\alpha$ & $\beta$ & $\nu_\perp$ & $\nu_\parallel$ & $z$ \\ \hline
DCMM\cite{Lubeck} & 0.141(24) & 0.382(19) & 1.347(91)& 1.87(13)& 1.393(37)   \\
DCMM* & 0.159(3) & $<$ 0.31 & 1.095(5)& 1.75(5) & 1.51(5) \\
CCMM*  & 0.1596(2) & 0.277(18) & 1.096(4)& 1.74(1) & 1.52(1) \\ \hline
DP~\cite{Henkel} & 0.1594 & 0.2764 & 1.0969& 1.733 & 1.5807 \\ \hline
\end{tabular} \end{small} \end{center} \vspace{-3mm}
\caption{ Estimates of the critical exponents in the literature and the present work (*) compared to DP.}
\label{table:I}
\end{table}
%========================================================

\paragraph{Continuous variant of the model:}
The discrete dynamics of the DCMM has several shortcomings. On the one hand, in finite systems the control parameter $\phi$ is quantized in steps of $1/L$, leading to systematic errors which have not been taken into account. On the other hand, the background noise is difficult to characterize because of its telegraphic nature. For this reason we define a different variant of the conserved Manna model, where the background is modeled by a continuous variable. We find that this continuous conserved Manna model (CCMM) exhibits a particularly clean DP scaling. 

The CCMM is defined on a one-dimensional lattice with $L$ sites and periodic boundary conditions, where we associate with each site $j$ a real-valued variable $E_j$ called energy. A site is declared as active if $E_j\ge 1$. In each time step all active sites are synchronously updated by redistributing the fractions $\xi_j$ and $1-\xi_j$ of the energy $E_j$ to the two neighboring sites, where $\xi_j\in(0,1)$ are random numbers. Clearly this update rule conserves the total energy $E= \sum_j E_j$. The model  
exhibits an absorbing phase transition when the energy density $e= E/L$ crosses a certain threshold value $e_c$. 

%========================================================
\begin{figure}[t]
\centering
\includegraphics[width=9cm]{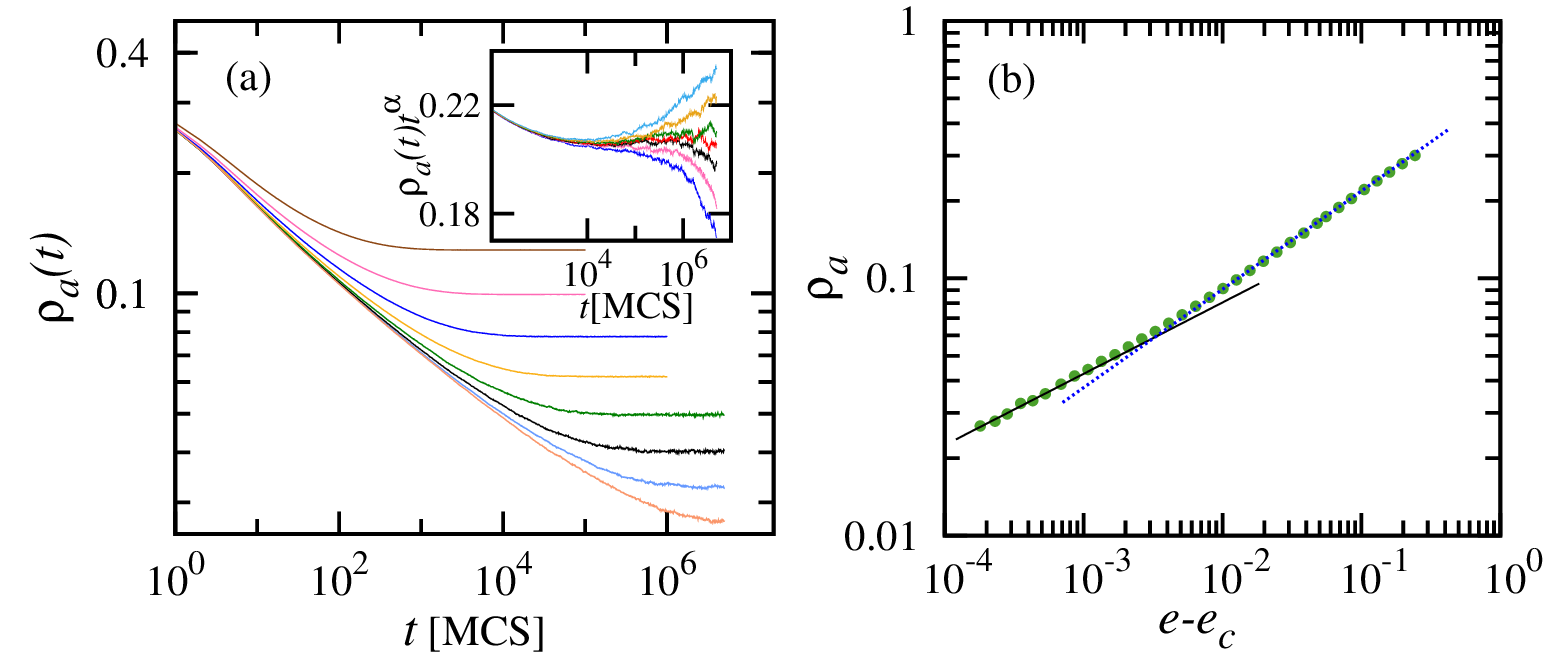}
\includegraphics[width=9cm]{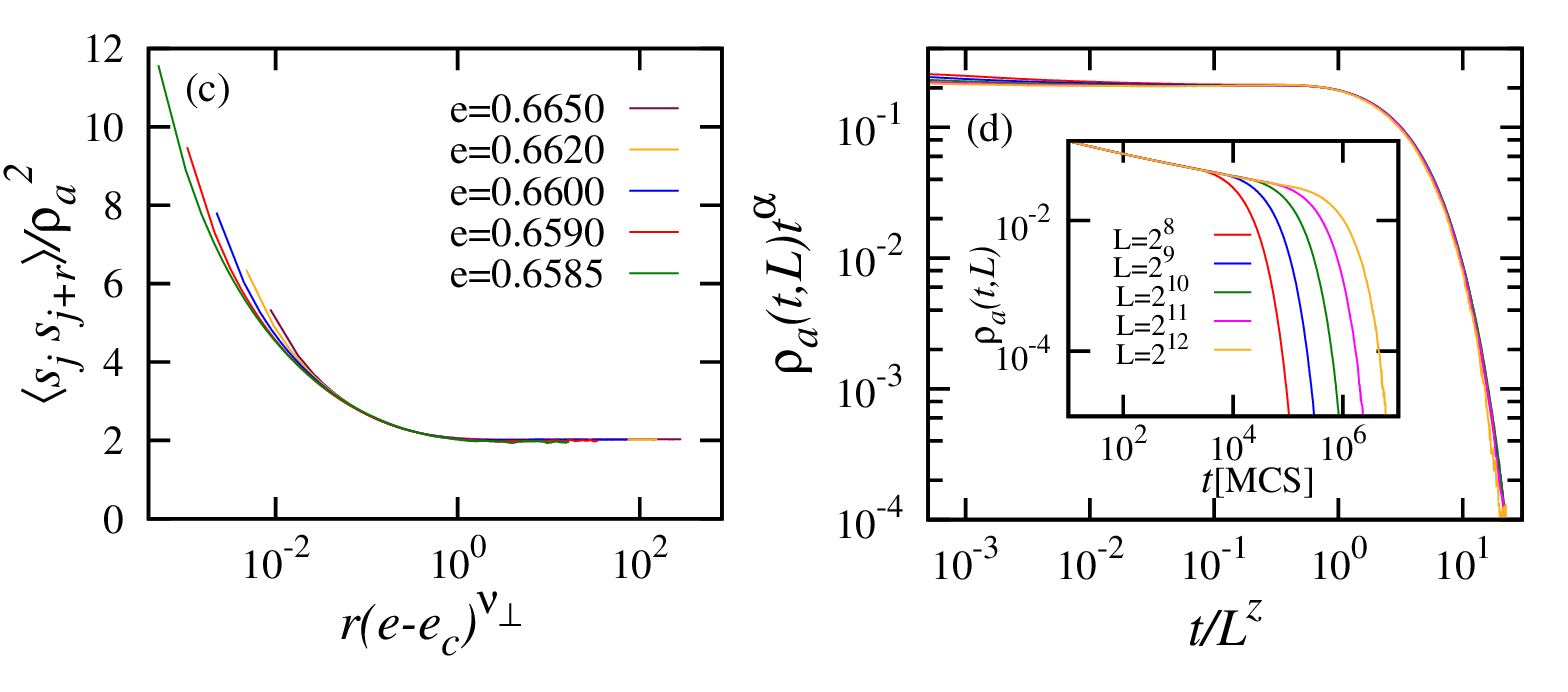}
\vspace{-9mm}
\caption{ CCMM: (a)  Search for the critical point using natural initial conditions (inset) and saturation in the supercritical phase in a system with $L=10^5$ sites. (b) Stationary density of active sites as a function of $\Delta=e-e_c$. (c-d) Data collapses for spatial correlations and finite size effects analogous to Fig.~\ref{fig:fig4}.}
\label{fig:fig5}
\end{figure}
%========================================================

\paragraph{Numerical results for the CCMM:}
For randomly distributed energies the continuous model shows the same undershooting of $\rho_a(t)$ as the discrete model. This can be avoided by using natural initial states [see Fig. \ref{fig:fig5}(a)]. Searching for the transition point (see inset) we find the critical energy $e_c=0.65797(1)$ and the decay exponent $\alpha= 0.1596(2)$. Repeating the above analysis in the stationary state $\Delta=e-e_c>0$, we find $\beta = 0.277(18)$ [solid line in Fig. \ref{fig:fig5}(b)] while farther away from criticality the local slope increases towards $0.381$  (dashed line) compatible with earlier results~\cite{Lubeck}. Moreover, the data collapse of $\rho_a(t)/\rho_a$ against $t\Delta^{\nu_\parallel}$ (not shown) gives the estimate $\nu_\parallel=1.74(1)$, while the data collapse for the two-point correlation function in Fig.~\ref{fig:fig5}(c) gives an estimate $\nu_\perp = 1.096(4).$ A finite-size scaling analysis leads to the dynamical exponent $z=1.52(1)$ [see Fig. \ref{fig:fig5}(d)].  Both $\nu_\perp$ and $\nu_\parallel$ are in good agreement with the corresponding DP values. 

\begin{figure}[t]
\centering
\includegraphics[width=86mm]{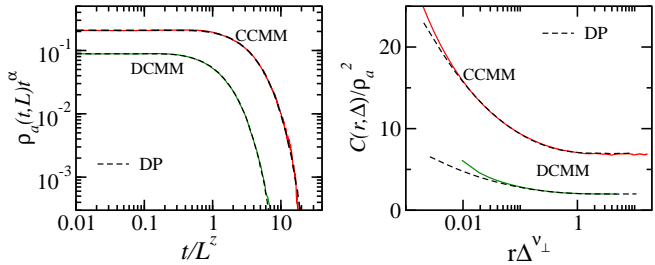}
\vspace*{-8 mm}
\caption{ Left: Finite-size data collapse for the DCMM and the CCMM (shifted) compared with the finite-size scaling function of DP (dashed black line).  Right: Analogous comparison for the stationary correlation function.}\label{fig:fig6}
\end{figure}

\paragraph{Comparison of scaling functions:}
In Fig.~\ref{fig:fig6} we overlay the data collapses of Figs.~\ref{fig:fig4} and~\ref{fig:fig5} with the appropriately shifted scaling functions $\cal F$ and $\cal G$ of DP (black dashed lines) which were determined numerically in a directed bond percolation process. As can be seen, the curves coincide almost perfectly for both the CCMM and the DCMM. A similar coincidence for the off critical  scaling function can be found in the Supplemental Material~\cite{SM}.

%============================
%============================

\begin{figure}[t]
\centering
\includegraphics[width=42mm]{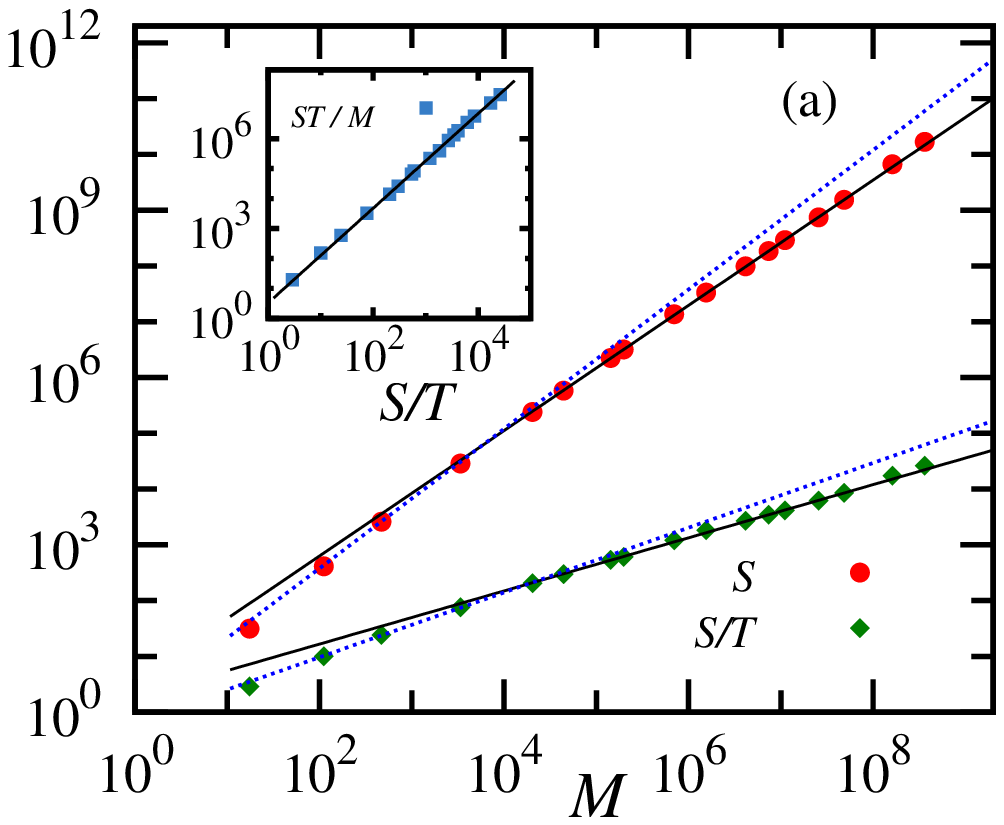}
\includegraphics[width=42mm]{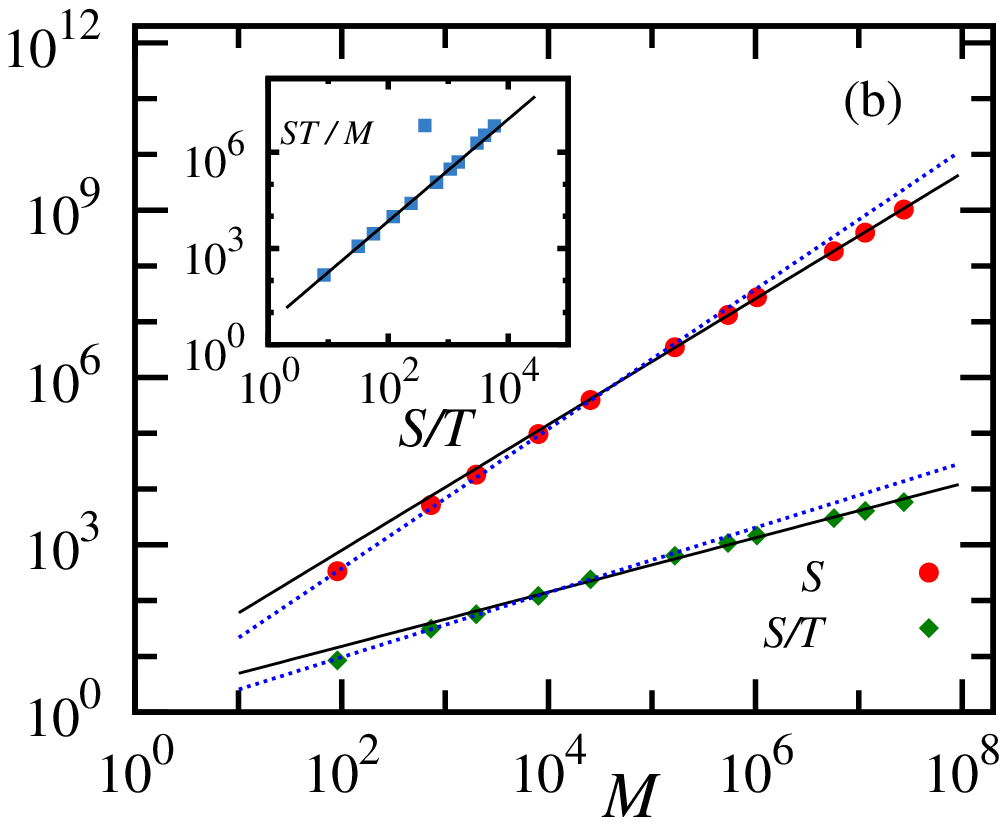}
\vspace{-3mm}
\caption{ Seed simulations in the CCMM (a) and DCMM (b) with $L=10^5$ sites. The main figure shows  both $S$  and $\frac S T$ versus $M$ together with the best fit over four decades (solid lines). The  inset shows $\frac{ST}{M}\sim \left( \frac{S}{T}\right)^z,$ 
fitted over three decades (solid lines). The dashed lines indicate  previously measured values~\cite{Lubeck}.}
\label{fig:fig7}
\end{figure}

\paragraph{Seed simulations:}
In seed simulations, we first let the system evolve  at a given $\Delta<0$ until it reaches a natural absorbing configuration~\cite{Natural}. In the discrete (continuous) model we then choose a random site $j$ with $n_j>0$ ($E_j>0$) and transfer  a particle (energy portions $E_k/2$) from randomly chosen  site(s) $k$ to the target site $j$ until it becomes active. This procedure conserves the number of particles (energy) without destroying the background correlations. We then measure the mean mass $M$, the mean survival time $T$ and the mean area $S$ of the generated clusters. In the stationary state near criticality, these quantities are expected to scale as
\begin{equation}
M\sim |\Delta|^{-\gamma}, \quad  T\sim |\Delta|^{-\tau}, \quad S\sim |\Delta|^{-\sigma} \,,
\label{MTS}
\end{equation}
where $\gamma= \nu_\perp + \nu_\parallel -2\beta$,  $\tau = \nu_\parallel -\beta$, and $\sigma= \nu_\perp + \nu_\parallel -\beta$. Although we could not reproduce DP exponents in these scaling laws individually, it turns out that $\Delta$-independent ratios of these quantities do exhibit a clean DP scaling. For example, Eq.~(\ref{MTS}) implies the relations
\begin{equation}
 S\sim   M^{1+\frac{ \beta}{\gamma} }\,,\quad
S/T  \sim M^{\frac{ \nu_\perp}{\gamma}}\,, \quad
ST/M\sim \left( S/T\right)^z \,. 
\end{equation}
As shown in Fig.~\ref{fig:fig7}, the  exponents  $\beta/\gamma$, $\nu_\perp/\gamma$  and  $z$ are estimated by $0.127(7), 0.486(9), 1.591(20)$ for DCMM, and by  $0.123(3), 0.476(5), 1.579(50)$ for CCMM respectively. Using the scaling relations $\gamma= \nu_\perp + \nu_\parallel -2\beta$ and $\nu_\parallel= \beta/\alpha$ these estimates are consistent with those of Table. \ref{table:I} and in excellent agreement with the corresponding DP values 
$\beta/\gamma=0.1214$, $\nu_\perp/\gamma=0.4816$, and $z=1.5807$.

%========================================================
%========================================================
\paragraph{Conclusions:} 
We have shown that the dynamics of the conserved Manna model  in (1+1) dimensions gradually removes the disorder in the background by itself. As a consequence, the problem of undershooting can be resolved by using homogeneously active initial states with natural background correlations. This leads to a slightly different estimate of the critical density and therewith to different critical exponents which then turn out to be mostly compatible with those of DP.  As shown in the Supplemental Material~\cite{SM}, analogous results were obtained in other  sister models belonging to Manna  class, namely in
the  (1+1)-dimensional CTTP and the CLG on a ladder.
 All our findings suggest that the Manna class is not independent but rather an extension or perturbation of DP with nontrivial boundary effects~\cite{Bonachela2}.

\end{document}

\title{Supplemental Material for \\ ``Fixed-Energy Sandpiles Belong Generically to Directed Percolation''}
\author{Mahashweta Basu$^1$, Urna Basu$^1$, Sourish Bondyopadhyay$^1$, P. K. Mohanty$^1$, and Haye Hinrichsen$^2$}
\affiliation{$^1$TCMP Division, Saha Institute of Nuclear Physics,
1/AF Bidhan Nagar, Kolkata 700064, India.\\
$^2$Universit\"at W\"urzburg, Fakult\"at f\"ur  Physik  und Astronomie, 97074  W\"urzburg, Germany.}

\begin{abstract}
In this supplement we provide additional details of the numerical simulations and further arguments to support our conjecture that the Manna class should be related to DP. We discuss the dynamics of the background field, the influence of the initial state, and  explain why previous works may have underestimated the critical threshold. Moreover, we present further numerical data for the conserved threshold transfer process (CTTP) and the conserved lattice gas (CLG) on a ladder.
\end{abstract}

\maketitle
\parskip 2mm

%========================================================================================
\section*{Motivation to doubt the existence\\ of an independent Manna class.}
%========================================================================================
%
In our opinion the following observations indicate that the commonly accepted scenario of an independent Manna class is probably not correct:

\begin{itemize}
\item
All existing evidence for an independent Manna class is numerical. 

\item 
While for all other established universality classes of absorbing phase transitions the numerical estimates are stable in two or more digits, the estimates for the Manna class are scattered over a wide range. For example, the exponent $\alpha=0.141(24)$ quoted in \cite{Lubeck} comes with an error of 17\%, which is unusual.

\item
Seed simulations are notoriously difficult to perform and are plagued by the same initial-state-dependence as other models with infinitely many absorbing states~\cite{JensenDickman93}.

\item
The Manna class and DP are known to have the same mean field theory and the same upper critical dimension $d_c=4$~\cite{LubeckHucht01,LubeckPRL,LubeckHeger03b}. 

\item
For homogeneous initial states the density of active sites in a critical system is expected to decay as $\rho_a(t) \sim t^{-\alpha }$ while in seed simulations the survival probability decays as $P_s(t) \sim t^{-\delta }$. In the Manna class the exponents $\alpha=\beta/\nu_\parallel$ and $\delta=\beta'/\nu_\parallel$ are reported to be different. On the other hand, one finds $\beta=\beta'$ within error bars~\cite{LubeckHeger03b}, which is a contradiction in itself. Contrarily Lee and Lee find that the scaling relations hold but then there is a mismatch in the exponent $\nu_\perp$~\cite{Lee3}.

\item
Some of the critical exponents for the Manna model and the CTTP~\cite{LubeckHeger03b} were found to be different. Although such a \textit{splitting of universality} may be attributed to the more deterministic character of the updates in the CTTP~\cite{Lee2}, it is still puzzling that only some exponents ($\beta'$,$\nu_\perp$, $\nu_\parallel$) are different while others (e.g. $\beta$, $\alpha$, $\delta$, $z$) still coincide in both cases.

\end{itemize}

On the other hand, Bonachela and Mu\~noz~\cite{Bonachela2} have convincingly demonstrated that boundary effects of models in the assumed Manna class are distinct from those of DP. Any questioning of the Manna class must be compatible with their findings. We will come back to this point at the end of this supplement.

%========================================================================================
\section*{Dynamics of the background field}
%========================================================================================

All models, which are believed to belong to the Manna class (MC), can be interpreted as a DP-like spreading process coupled to a conserved background field (i.e. the particle density), which determines the local rate for offspring production. This interpretation was brought into a transparent form by Vespignani et al.~\cite{FES}, who suggested that the essential properties of phase transitions in the Manna class are captured on a coarse-grained level by a system of two coupled Langevin equations for an \textit{activity field} $\rho_a(x,t)$ and a \textit{background field} $\phi(x,t)$:
\begin{eqnarray}
\label{Langevin1}
\partial_t \rho_a &=& r \rho_a  - b \rho_a^{2} + \nabla^2 \rho_a+ \sigma {\sqrt \rho_a}\eta +\omega \rho_a \phi\, \label{eq:Langevin1}\,,\\
\partial_t \phi &=& D \nabla^2 \rho_a\label{eq:Langevin2}\,,
\end{eqnarray}
where $r, b, \sigma, D, \omega$ are constants and $\eta(x,t)$ denotes an uncorrelated Gaussian noise. The first equation is just the usual Langevin equation for directed percolation except for the term $\omega \rho_a \phi$, in which the usual fixed rate for offspring production is now replaced by the background field $\phi(x,t)$. The second equation describes how the background field is reordered in the presence of activity. One can easily verify that these equations conserve the total number of particles $\int dx \,\phi(x,t)$.

Our work is motivated by the conjecture that the spreading process in the first equation evolves \textit{superdiffusively} with a dynamical exponent $z<2$ while the conserved background evolves only \textit{diffusively}, providing effectively a disordered background on which the spreading process takes place. The spreading process in turn reorders the background, but on a much slower time scale. Therefore, depending on the initial state, the disorder in the background may have an influence on the temporal evolution over a long time. If we use reactived steady state configurations as the initial states,  here referred to as natural initial conditions, the background disorder is assured to be low and the critical behavior of the model is that of directed percolation. This  suggests that an independent Manna class does not exist, rather fixed energy sandpiles may be viewed as DP-processes running on a self-flattening background.

%========================================================================================
\section*{Initial state dependence}
%========================================================================================

\subsection*{Undershooting}

As demonstrated in Fig. 1 of the Letter, a strongly disordered background leads to the phenomenon of \textit{undershooting}. Since the dynamics of the background is slow compared to the spreading process, the situation can be compared with a DP process in presence of spatially quenched disorder~\cite{Moreira96,Janssen97,CafieroEtAl98,Webman98,Vojta04,VojtaDickison05,VojtaLee06,DahmenEtAl07}, but with the important difference that in the present case the disorder is not frozen but slowly modified by the process itself. This causes a glassy behavior which is known to slow down the decay of the density of active sites $\rho_a$. This makes it plausible why the estimates for the exponent $\alpha=0.141(24)$ quoted in~\cite{Lubeck}, describing the decay $\rho_a(t)\sim t^{-\alpha}$, tends to underestimate the DP exponent $\alpha \approx 0.159$ and why the reported error bar is so large. 

Although the process gradually homogenizes the background, this happens on a much slower time scale so that a supercritical system first reaches a quasi-stationary state, where the disorder of the background is still present on scales above the correlation length. Because of the expected non-linear response $\rho_a(x) \sim (\phi(x)-\phi_c)^\beta$ with $\beta<1$ this disorder will on average \textit{decrease} the quasi-stationary density of active sites, leading to the observed undershooting. However, as time proceeds the disorder in the background field slowly disappears so that $\rho_a(t)$ increases again and eventually reaches its `true' asymptotic value.  This rebound takes more than two orders of magnitude in the simulation time, which -- if not properly taken into account -- may have led to additional errors in previous steady-state simulations.

\subsection*{Explaining undershooting: \\Spreading of localized initial states}

\begin{figure}[t]
\centering \includegraphics[width=85mm]{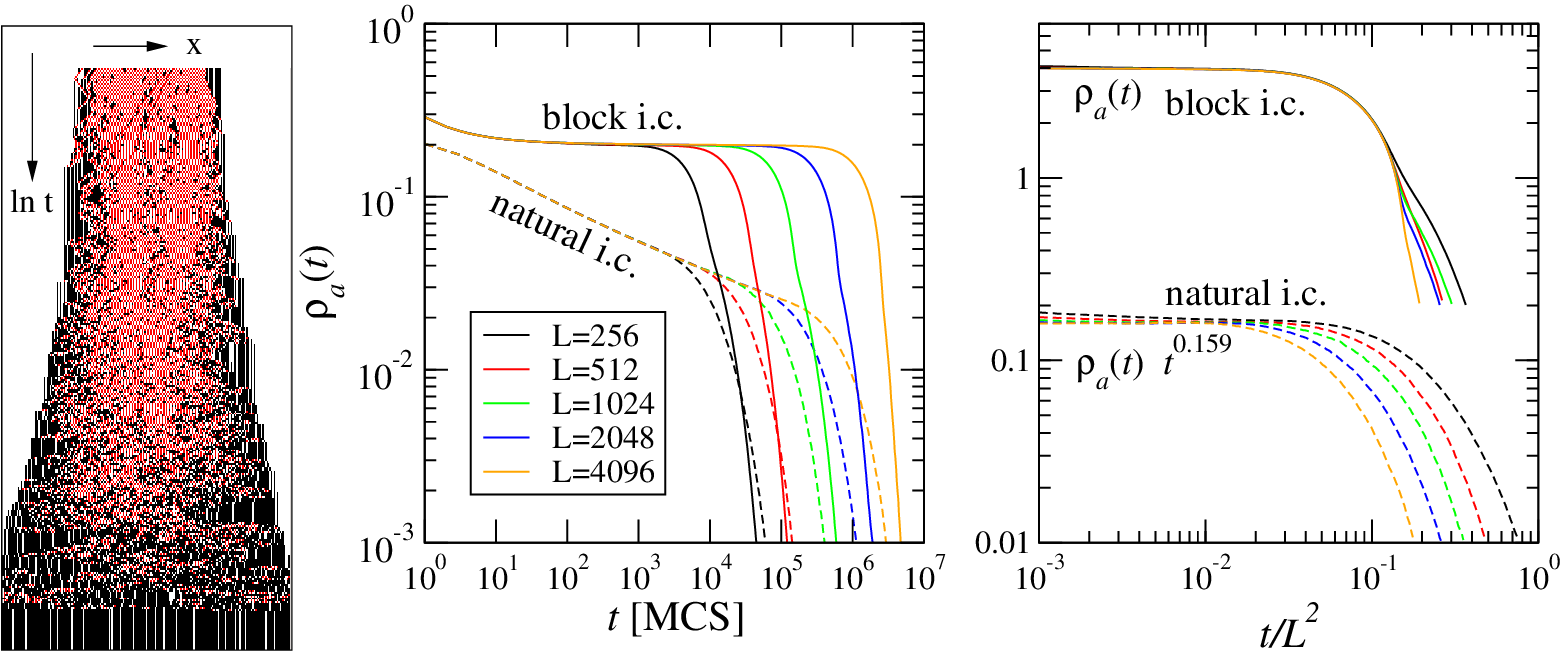}
\caption{\label{fig:hump} Cloggy initial state in the CMM at the critical density $\phi=\phi_c$. Left: If all particles are initially concentrated in a single hump this region will continue to spread until the ends touch each other. Middle: Density of active sites as a function of time for various system sizes for cloggy (solid lines) and natural initial conditions (dashed lines). Right: Unlike natural initial states, which scale with the DP exponent $z\approx 1.58 < 2$, the curves for cloggy initial conditions collapse if they are plotted against $t/L^2$. This confirms that the background field spreads diffusively (see text). }
\end{figure}

The phenomenon of undershooting can also be explained by demonstrating the diffusive character of the background dynamics  in the case of highly disordered initial states \footnote{We are indebted to referee A for suggesting this explanation.}. To see this let us consider a ``cloggy'' initial state at the critical density $\phi=\phi_c$, where all particles are concentrated in a single hump. For example, in the DCMM we could pack $N=\phi_c L$ particles in a single dense block by placing $n_j=2$ particles on all sites $j=0 \ldots N/2$ while keeping the rest of the system empty. Clearly, the process inside the block is supercritical and thus the particles will begin to spread, supporting a high level of local activity. The spreading region is expected to grow until the two ends meet each other because of periodic boundary conditions. At this moment the density inside the block, which now covers the whole system, becomes critical, and therefore the density of active sites $\rho_a(t)$ is expected to break down. As demonstrated in Fig.~\ref{fig:hump}, the typical time scale for this density breakdown grows \textit{quadratically} with the system size $L$. Therefore, this numerical experiment confirms that the reordering of the background field, as described by Eq.~(\ref{eq:Langevin2}), is in fact diffusive.

%============================
%============================

The implication would be that fixed energy sandpiles starting with slightly inhomogeneous initial states at the critical point are characterized by \textit{two competing length scales} 
\begin{equation}
\xi_1(t) \sim t^{1/z} \quad \mbox{and} \quad \xi_2(t) \sim \sqrt{t}\,. 
\end{equation}
The larger length scale $\xi_1(t)$ is the usual DP-like correlation length of the cluster of active sites while $\xi_2(t)$ is the scale on which the inhomogeneities have been flattened out. Above criticality, $\xi_1(t)$ grows until it reaches a stationary value $\xi_1\sim(\phi-\phi_c)^{-\nu_\perp}$. At this moment the activity is still concentrated in many small humps, giving on average a smaller activity. However, the humps continue to spread diffusively until they touch each other, leading to a subsequent increase of the average  activity density. The `true' asymptotic density is reached only when all humps have been flattened out.

\subsection*{Comparing various types of initial states}

Having identified undershooting and related memory effects as one of the main reasons for the numerical difficulties in the conserved Manna model with homogeneous initial conditions, it is near at hand to create initial states for which the disorder in the background field is low. 

In the Letter we favor \textit{natural} initial conditions by letting the system first evolve into a stationary or absorbing configuration and then {\it reactivating } it by diffusion for a single time step\footnote{Note that the critical density in the DCMM of about 0.9 is rather high so that diffusion for a single Monte Carlo sweep generates already a high density of active sites.}. There are, of course, several other ways to create low-disorder initial configurations. For the DCMM we may, for example, create
\begin{itemize}
\item 
\textbf{flat absorbing states,} where vacant sites are distributed deterministically as equidistant as possible by setting $n_{\lfloor i/\phi \rfloor}=1$ for $i=0,1\ldots \lfloor \phi L \rfloor$ and zero otherwise. This state has an artificially flat profile.
\item 
\textbf{random absorbing states}\footnote{We thank referee A for this suggestion.}, by first setting $n_i=1$ with probability $\phi$ and zero otherwise, and then adding (deleting) particles  at vacant (occupied) sites so that the total number of particles is $\lfloor \phi L \rfloor$. This state is not flat but compared to random active states the disorder is significantly reduced. 
\end{itemize}
In both cases the states are then reactivated by diffusion for a single Monte-Carlo sweep.

%============================
\begin{figure}[t]
\centering
\includegraphics[width=87mm]{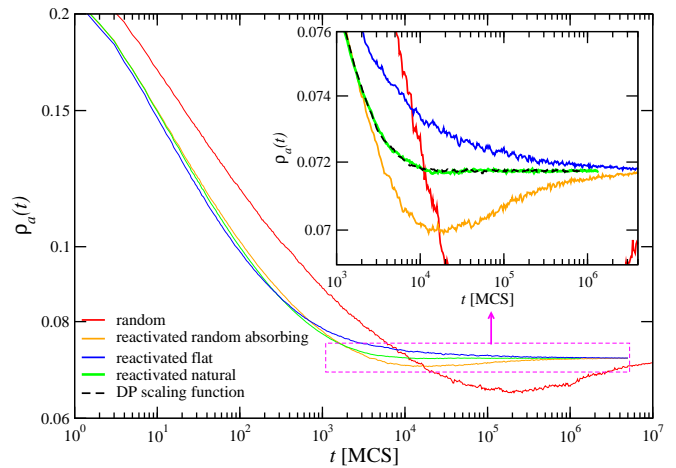}
\caption{\label{fig:compare} Comparison of the decay for different initial states in DCMM with $\phi-\phi_c=0.0128$. The inset shows a zoom of the same data. The expected DP scaling function $\cal R$ (see text), shown here as black dashed line, compares well with the reactivated natural initial condition.
}
\end{figure}
%============================

As shown in Fig.~\ref{fig:compare}, the standard random initial conditions discussed in the Letter (red curve) lead to a pronounced undershooting, while the other curves appear to be quite similar. However, as can be seen in the inset, reactivated random absorbing states (orange curve) still undershoot a little, meaning that the disorder in the background field is still strong enough to delay the dynamics. On the other hand, natural and flat initial conditions do not undershoot. This demonstrates that there are many ways to engineer an initial state in such a way that undershooting is suppressed.

We favor \textit{natural} initial conditions for two reasons. On the one hand, there has been a similar debate in the context of seed simulations, where it turned out that a naturally correlated background yields the best numerical results~\cite{JensenDickman93}. On the other hand, the data shown in the figure is expected to obey the scaling form
\begin{equation}
\label{G}
\rho_a(t,\Delta) = t^{-\alpha} {\cal R}(t \Delta^{\nu_\parallel})\,,
\end{equation}
where $\Delta=\phi-\phi_c$ is the distance from criticality and $\cal R$ is a universal scaling function which determines the form of the curve in the figure. If, as suggested in the Letter, the DCMM model indeed belongs to DP, then this scaling function should be the same as in DP. In the inset of Fig.~\ref{fig:compare} we overlaid the scaling function of ordinary directed bond DP (black dashed line). As can be seen, natural homogeneous initial conditions lead to a perfect coincidence. This provides a third example for the coincidence with DP scaling functions, in addition to the finite-size and two-point correlation scaling functions discussed in the Letter.   The  perfect  matching of  $\rho_a(t)$ for natural initial condition with the scaling function of DP also confirms that the reactivation process for a single time step does not destroy the  natural correlations of the stationary state.

\subsection*{Determination of the critical density}

We suspect that the phenomenon of undershooting may have led to a systematic underestimation of critical densities in previous works. As an example, we show that the commonly accepted critical point of the DCMM $\phi_c=0.89199(5)$ quoted in~\cite{Lubeck} is incorrect and has to be replaced by $0.89236(3)$ -- a value far beyond the previous error bars. 

\begin{figure}[t]
\centering
\includegraphics[width=87mm]{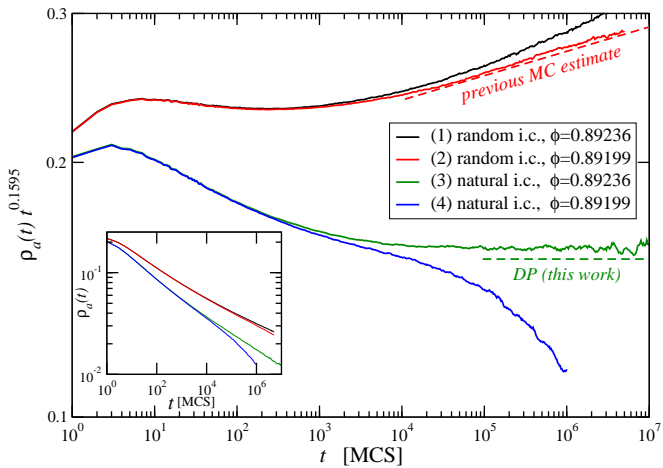}
\caption{\label{fig:decay} Temporal decay of the   activity density in the DCMM for various types of initial states. The inset shows the raw data while the main figure displays the same data divided by the expected power law of DP.}
\end{figure}

In Fig.~\ref{fig:decay} we illustrate step by step how it comes to such a discrepancy:

\begin{itemize}
\item[(1)] 
If we would simulate the process with random initial conditions at the correct critical point $\phi_c=0.89236(3)$ the glassy influence of the randomness slows down the dynamics and leads to a soft positive curvature of the data (black curve in Fig.~\ref{fig:decay}) which looks as if the process was slightly supercritical. 
\item[(2)] 
This observation may have tempted previous authors to `compensate' this curvature by slightly lowering the estimate of $\phi_c$. A good-looking ``power law'' $\rho_a\sim t^{-\alpha}$ (red curve) is then obtained for $\phi_c=0.89199(5)$. The corresponding slope (red dashed line) is roughly $-0.14$, compatible with previously reported estimates of $\alpha$.
\item[(3)] 
However, if we simulate the process with natural initial conditions for this value for $\phi_c$ we would obtain the blue curve which is clearly subcritical.
\item[(4)] 
Searching for a straight line with natural initial conditions (green curve) one obtains the corrected estimate  $\phi_c=0.89236(3)$ and a power law decay $\rho_a(t)\sim t^{-\alpha}$ with the exponent $\alpha=0.159(3)$ which is now compatible with DP.
\end{itemize}

%============================
%============================

\begin{figure}[t]
\centering
\includegraphics[width=87mm]{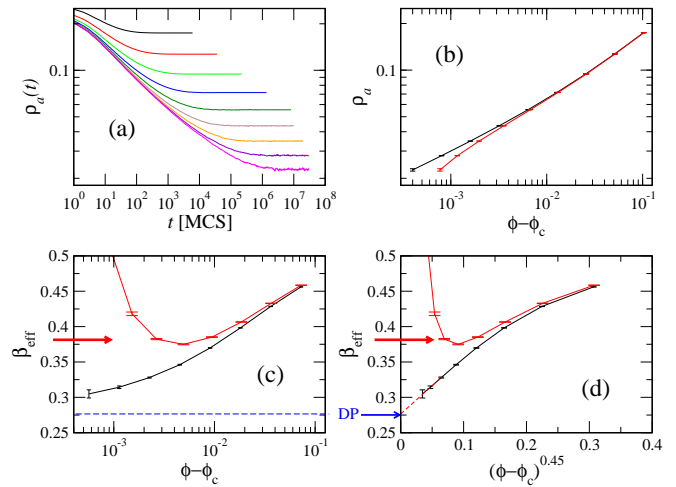}
\caption{\label{fig:wrong-phic} Estimation of the static exponent $\beta$. This figure illustrates how the data of Fig. 3 in the Letter (black lines) would look like when using the literature value $\phi_c=0.89199(5)$ (red lines) instead of the corrected estimate  $\phi_c=0.89236(3)$.}
\end{figure}

\noindent
To support this point further, we demonstrate in Fig.~\ref{fig:wrong-phic} how the data analysis for the stationary state would look like if we used instead of the corrected estimate the old literature value  $\phi_c=0.89199(5)$. Already the raw data in panel (b) would exhibit a changing sign in the curvature, which has not been seen so far in absorbing phase transitions. Computing the local slopes in panel (c) one can easily see where the old estimate $\beta=0.382(19)$ in Ref.~\cite{Lubeck} comes from (marked by the red arrow), but with increasing numerical effort the effective exponent turns out to grow rapidly as we approach criticality, - a clear indication that the estimate of the critical point is incorrect.

%============================
%============================

%========================================================================================
\section*{Similar results in other models}
%========================================================================================

\subsection*{Conserved Threshold Transfer Process (CTTP)}

To further back up our numerical results, we investigate here the so-called Conserved Threshold Transfer Process introduced by Rossi \textit{et al}.~\cite{Rossi}, which is believed to belong to the Manna class. The CTTP is defined on a hypercubic lattice, where each sites can be empty or occupied by one or two particles. Sites with two particles are active. The model evolves by random-sequential updates, i.e. an active site is randomly selected and the two particles are independently moved to a randomly selected nearest neighbor, provided that there are less than two particles at the target site. The CTTP may be interpreted as a fixed-energy Manna sandpile with a height restriction, where the occupation of any site can not  exceed a value $n^*=2$.

While Rossi \textit{et al}. focused on the two-dimensional case, several authors have studied the CTTP in one dimension~\cite{LubeckCTTP,DickmanCTTP,Lee4}, obtaining contradicting results (see Table~\ref{tab:cttp}). In particular, all authors report different critical densities, indicating that they may have been using different variants of the model. Moreover, the critical exponents obtained in these studies differ significantly.

%====================CTTP=========================
\begin{table*}
\begin{center}
\begin{tabular}{|l|c|ccccc|} \hline\hline
Ref. & $\phi_c$ & $\alpha$ & $\beta$ & $\nu_\parallel$ & $\nu_\perp$ & $z$\\ \hline
L\"ubeck~\cite{Lubeck, LubeckCTTP} 	
& \quad$0.96929(3)\quad$ & $0.141$ & $0.382$ & $2.452$ & $1.760$ & $1.393$\\
Dickman \textit{et. al.}~\cite{DickmanCTTP}
& $0.92965(3)$ & - & $0.412$ & $2.41$ & $1.66$ & $1.45$\\
Lee~\cite{Lee4}
& $0.98285(5)$ & $0.118$ & $0.396$ & $3.36$ & $2.26$ & $1.49$\\
Dickman~\cite{DickmanPreprint}
& $0.92978(2)$ & $0.141$ & $0.289$ & $2.03$ & $1.36$ & $1.50$\\
this work 			
& $0.929735(15)$ & \; $0.155(5)$\; & \;$0.308(2)$\; & \;$1.74(1)$\; & \;$1.13(1)$\; & \;$1.52(2)$\, 
\\ \hline DP &  & $0.159$ & $0.277$ &  $1.733$ & $1.096$  & $1.580$ 
\\ \hline \hline
\end{tabular}
\caption{\label{tab:cttp} Reported estimates for the critical exponents of the one-dimensional CTTP compared with DP. The values have been rounded to three decimals.}
\end{center}
\vspace{-3mm}
\end{table*}
%=============================================

%============================
\begin{figure}[t]
\centering
\includegraphics[width=87mm]{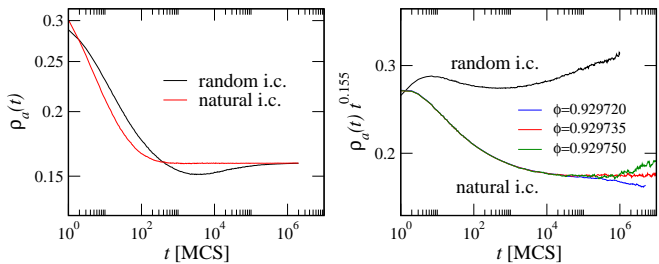}
\caption{\label{fig:cttp} CTTP in one dimension. The left figure confirms the phenomenon of undershooting for random initial conditions in the supercritical phase for $\Delta=\phi-\phi_c=0.0512$ on a chain with $L=2^{16}$ sites. The right panel demonstrates the pronounced difference between random and natural initial conditions. The data is divided by the expected DP power law. }
\end{figure}
%============================

In this context it is worthwhile to mention the work by R. Dickman in 2006~\cite{DickmanPreprint} as a remarkable contribution. In our opinion  he is the first one to fully realize the numerical subtleties in simulations of fixed energy sandpiles and their importance in determining the critical point of the CTTP. To overcome these difficulties he develops various empirical methods to compensate the curvature in the data. This leads to exponents which, in his words, ``\textit{are not very different from those of DP}'' (see Table~\ref{tab:cttp}). He points out that Kockelkoren and Chat\'e~\cite{Kockelkoren} as well as Ramasco \textit{et al}. ~\cite{Ramasco} also obtained numerical results close to DP in a related model (e.g. $\beta=0.29(2)$ compared to $\beta_{DP}=0.277$). Despite this obvious proximity to 
DP he was still convinced that fixed energy sandpiles should constitute independent universality class as the value of the decay exponent $\alpha=\beta/\nu_\parallel$ would be clearly different. Here we argue that this last remaining discrepancy in $\alpha$ can be resolved by using natural initial states.

\begin{figure}[t]
 \centering
 \includegraphics[width=87 mm]{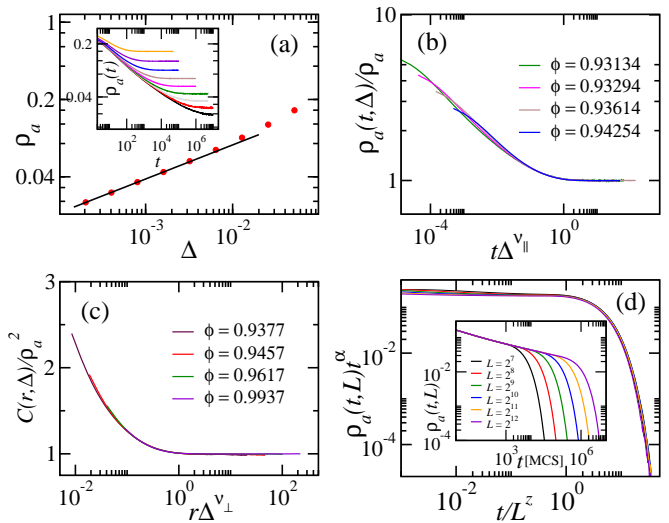}
 % sat.eps: 0x0 pixel, 300dpi, 0.00x0.00 cm, bb=(atend)
\caption{Conserved threshold transfer process (CTTP). (a) Stationary density of active sites as a function of $\Delta=\phi-\phi_c$. Inset: Saturation in the supercritical phase in a system with $L=10^5$ sites. (b) Data collapse according to Eq. (\ref{eq:nuparallel}) giving $\nu_\parallel=1.74(1)$.  (c) Data collapse for spatial correlations according to Eq.(\ref{eq:nuperp}) for different values of $\Delta$. (d) Finite-size scaling with $L=2^7,\ldots, 2^{12}$. The best collapse is obtained for $z=1.52(2)$.}
 \label{fig:cttp_beta}
\end{figure}

Redoing parts of the simulations  for CTTP \cite{DickmanPreprint} using natural initial conditions we estimate the critical density $\phi_c=0.929735(15)$, which is close to the value obtained by Dickman \cite{DickmanPreprint} by different correction methods. Again we observe the phenomenon of undershooting, which disappears when natural initial conditions are used (see Fig.~\ref{fig:cttp}). At the critical point $\rho_a(t)$ decays as $ t^{-\alpha}$  with $\alpha=0.155(5)$, in  excellent agreement with DP values. The saturation values of activity in the supercritical regime show a pronounced curvature when  plotted against $\phi-\phi_c$  in log scale [Fig. \ref{fig:cttp_beta}(a)], as already observed by Dickman. A linear fit  near the critical point gives an  estimate of $\beta= 0.308(2).$ This is compatible with  similar results obtained in \cite{DickmanPreprint} using finite size scaling.  Both values of $\beta$ are  not far from $\beta_{DP}$. 

 We complement these studies by a collapse of the curves $\rho_a(t,\Delta)$ in the supercritical phase for different $\Delta= \phi-\phi_c >0$ according to the scaling form
\begin{equation}
\rho_a(t,\Delta) = \rho_a {\cal R}(t\Delta^{\nu_\parallel}).\label{eq:nuparallel}
\end{equation}
As shown in Fig. \ref{fig:cttp_beta}(b), the best collapse is obtained for  $\nu_\parallel=1.75(1).$ Moreover, a data collapse of the spatial correlation function $C(r,\Delta)$  for different $\Delta$ according to the scaling form 
\begin{equation}
C(r,\Delta) =\rho_a^2 {\cal G}( r\Delta^{\nu_\perp})\label{eq:nuperp}
\end{equation}
 yields $\nu_\perp =1.13(1)$ [see Fig. \ref{fig:cttp_beta}(c)].  Finally, a standard finite size scaling collapse shown in Fig. \ref{fig:cttp_beta}(d) gives an estimate of $z=1.52(2)$. Our estimates of  the  critical exponents are listed in Table ~\ref{tab:cttp} along with those obtained in earlier works. The use of natural initial conditions leads to a coincidence of the decay exponent $\alpha$ with DP and brings $\nu_\perp$ and $\nu_\parallel$ closer to the corresponding DP values. 

%============================
%============================

We note that the mysterious splitting of universality observed by L\"ubeck and Heger~\cite{LubeckHeger03b}, stating that the conserved Manna model and the CTTP in 1D differ in some but not all critical exponents, disappears when natural initial conditions are used. The same applies to apparent violations of scaling. 

The CTTP and the conserved Manna model are very similar. In fact, if the height restriction in CTTP is lifted by allowing  $n^* \to \infty$, $i.e.$  by taking  $n^* \sim  20$ for all practical purposes,  the dynamics near  the critical density becomes identical with that of the conserved Manna model. It is therefore not surprising that both models exhibit the same type of critical behavior.

\subsection*{Conserved Lattice Gas (CLG)}

Another popular model which is believed to represent the Manna class is the so-called Conserved Lattice Gas. The CLG is defined on a $d$-dimensional hypercubic lattice, where each site is either empty or occupied by a single particle ($n_i=0,1$). The particles  interact via nearest neighbor  repulsion. This means that a particle, which has at least one neighboring particle, tries to move to one of the randomly chosen vacant neighbors. If all neighboring sites are occupied the particle  is unable to move.  Thus, a particle is active if it has at least one neighboring particle and at least  one vacant neighboring site. 

The CLG was originally introduced by Rossi \textit{et al}.~\cite{Rossi} who used a parallel update scheme with an exclusion principle. Since this method depends on the sequence in which the sites are updated, L\"ubeck used in a later study random-sequential dynamics, i.e. an active site is randomly selected and the particle jumps to one of the neighboring vacant sites with equal probability.

%============================
\begin{figure}[b]
\centering
\includegraphics[width=85mm]{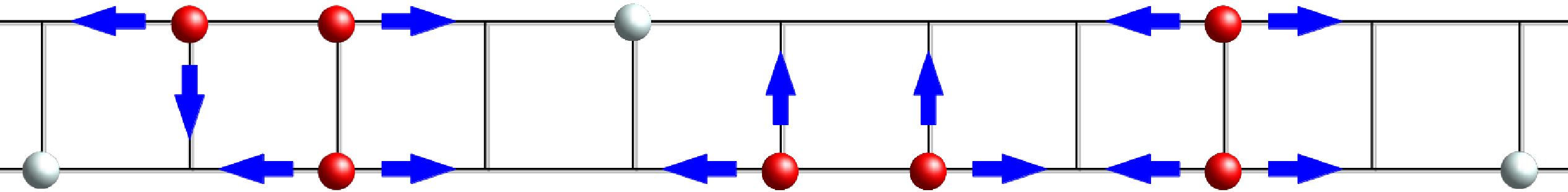}
\caption{Conserved Lattice Gas (CLG) on a ladder. Particles with a nearest neighbor are active (red bullets), otherwise they are inactive (gray bullets). During the update randomly selected active particles move to one of the empty nearest neighbors, as indicated by the arrows.}\label{fig:ladder}
\end{figure}
%============================

In one spatial dimension an active site has at most one neighboring vacant site, leading to deterministic hopping sequences of active particles. As a consequence the transition becomes  trivial in this case and is characterized by integer exponents~\cite{Lee1}. As a way out, Oliveira proposed to study the CLG on a ladder~\cite{OliveiraCLG1d}. Since  sites on a ladder have three neighbors each (see Fig.~\ref{fig:ladder}), some of the active  sites  have two vacant neighbors and  the   particle  from such a site hops {\it stochastically}\footnote{ Stochasticity in particle transfer is essential for a non-trivial 
critical behavior;  critical  behaviour of deterministic models  are known \cite{BTW} to be different.} 
as in DCMM or in  CTTP. However, the numerical analysis by Oliveira was restricted to the estimation of only one exponent, namely $\beta=0.40(1)$.

%============================
\begin{figure}[h]
\vspace*{0.8 cm}
\centering
\includegraphics[width=8.7 cm]{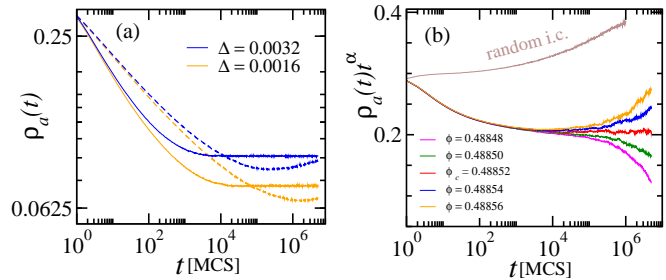} 
% ladder_rc.eps: 0x0 pixel, 300dpi, 0.00x0.00 cm, bb=(atend)
\caption{ CLG on a ladder :  (a) Comparison of decay of activity  from random and natural 
initial  configurations for  $\Delta=0.0016, 0.0032$ for a system of size $L=10^5$.
(b) Determination of critical point : $\rho_a(t) t^\alpha$  saturates  at  the critical point $\phi_c=0.48852(2) .$ The random initial condition shows  an   apparent supercritical feature at the critical point $\phi_c.$ }
 \label{fig:clg_rc}
\end{figure}
%============================

Following this idea, we  numerically analyze the CLG on a ladder  of linear size $L$, 
i.e. $2L$  sites  in  total. As demonstrated in Fig.~\ref{fig:clg_rc}(a), we observe again the phenomenon of undershooting in the decay  of $\rho_a(t)$ starting from a random configuration. This undershooting disappears if natural initial conditions are used. With natural initial conditions we obtain the critical point $\phi_c= 0.48852(2)$, which is a  bit higher compared to  the  earlier estimate 
$0.4755(2)$ reported in~\cite{OliveiraCLG1d}. 

\begin{figure}[t]
 \centering
\includegraphics[width=8.2 cm]{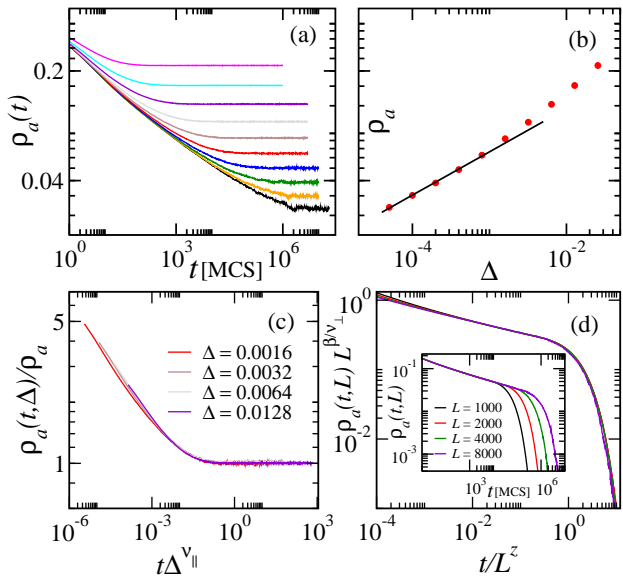}
%\vspace*{-2cm}
 % ladder_rc.eps: 0x0 pixel, 300dpi, 0.00x0.00 cm, bb=(atend)
 \caption{CLG on a ladder : (a) Saturation of activity in the supercritical regime for 
$\Delta=\phi-\phi_c= 0.00005, 0.0001, 0.0002,\dots 0.0256$ in a system of size $L=10^5.$ (b) Corresponding saturation values plotted against $\phi-\phi_c$; the solid line corresponds to the estimate
$\beta=0.275(6).$ (c) Determination of $\nu_\parallel$ from a data collapse following Eq. (\ref{eq:nuparallel}) using the saturation data for $\phi-\phi_c=0.0016,0.0032,0.0064$ and $0.0128$ from (a), giving the estimate $\nu_\parallel=1.76(2).$ (d)  A finite size scaling collapse for $L=1000, 2000, 4000, 8000$ yields $z=1.50(2)$ and $\beta/\nu_\perp=0.251(4)$; the inset shows the corresponding raw data.}
 \label{fig:clg_sat}
\end{figure}

At the critical point the   log scale plot of $\rho_a(t)$ gives the slope $\alpha= 0.1553(1)$, which is  in excellent  agreement with DP [see Fig.~\ref{fig:clg_rc}(b)]. In the supercritical regime, the activity density saturates at a steady value $\rho_a$,  which  varies as  $(\phi-\phi_c)^\beta$ with $\beta=0.275(6)$ [see figures Fig.~\ref{fig:clg_sat}(a)-(b)]. These saturation curves for different values of $\Delta =\phi-\phi_c$  can be collapsed as usual in accordance with  Eq. (\ref{eq:nuparallel}). This is demonstrated in Fig. \ref{fig:clg_sat}(c) for $\Delta= 0.0016,0.0032,0.0064,0.0128$ giving the estimate $\nu_\parallel=1.76(2).$ Likewise, finite size scaling collapse following $\rho_a(t,L) = L^{-\beta / \nu_\perp} \tilde{\cal F}(t/L^z),$ shown in Fig.~\ref{fig:clg_sat}(d),  gives  $z= 1.50(2) $  and $\beta/\nu_\perp=0.251(4).$

%==================CLG===========================
\begin{table}[b]
\begin{center}
\begin{tabular}{l|ccccc} \hline\hline
Ref. &  $\alpha$ & $\beta$ & $\nu_\parallel$ & $\beta/\nu_\perp$ & $z$\\ \hline
CLG 1D\cite{Lee1}
&  $ 1/4$ & $1$ & $4$ & $1$ & $2$\\
modified \cite{Oliveira2}
&  $0.13(1)$ & $0.277(3)$ & $2.41$ & $0.223(5)$ & -\\
ladder  \cite{OliveiraCLG1d}  	
&  - & $0.40(1)$ & - & - & -\\
this work 			
&  $0.1553(1)$ & $0.275(6)$ & $1.76(2)$ & $0.251(4)$ & $1.50(2)$ \\
 \hline DP &  $0.159$ & $0.277$ &  $1.73$ & $0.252$& $1.58$ 
\\ \hline \hline
\end{tabular}
\end{center}
\vspace{-3mm}
\caption{ Reported estimates for the critical exponents of 
CLG on a ladder   are compared  with DP.}\label{tab:clg}
\end{table}
%=============================================

A summary of these critical exponents can be found in Table \ref {tab:clg}, along with the results 
of previous studies  of CLG on a ladder  and  a  modified version of CLG in one dimension studied by Fiore and  Oliveira \cite{Oliveira2}, where both adjacent active particles  may  jump  simultaneously  to their respective neighboring  vacant sites.

%============================
%============================

%========================================================================================
\section*{Claim and outlook}
%========================================================================================

In our work we have shown in the example of the one-dimensional conserved Manna model that the commonly accepted critical point $\phi_c=0.89199(5)$ is incorrect. This means that all critical exponents, which were estimated on the basis of this critical point, have to be revisited.

Using \textit{natural initial conditions} we find a critical behavior which is the same as or at least very close to DP, leading us to the convincing but nevertheless speculative conjecture that it actually \textit{is} DP. This conjecture is based solely on numerical results and needs to be substantiated further by extensive high-performance simulations on supercomputers. 

Note that so far this DP-conjecture is restricted to FES models with periodic boundary conditions in one spatial dimension, where the conserved background field flattens by itself. Whenever this mechanism is disturbed, e.g. by other types of boundary conditions or the presence of external fields, the background field could play a non-trivial role and the resulting behavior will differ from DP. In particular the results obtained by Bonachela and Mu\~noz~\cite{Bonachela2}, who proposed an efficient method to distinguish DP from MC by boundary effects, remain valid. This means that FES models are still `more' than directed percolation in the sense that the conserved background leads under certain circumstances to additional features which cannot be seen in ordinary DP models. However, our results suggest that these models do not constitute an independent universality class with an autonomous set of critical exponents and scaling functions.

We note that the observation of a self-flattening background field seems to be incompatible with earlier field-theoretic findings by Pastor-Satorras and Vespignani~\cite{FieldTheory}, who pointed out that quartic vertices become relevant, driving the system away from DP. In this context it would be important to study whether the results of the present paper, which are restricted to one spatial dimension, can also be confirmed in higher dimensions.

Even if the Manna class turned out to be DP, the simple conclusion that ``SOC is just DP'' would be premature. On the one hand, driving and dissipation communicate via the background field, which could play a non-trivial role in such situations. On the other hand, the relationship between SOC and fixed energy sandpiles as such is still debated in the literature~\cite{Fey1,Fey2,Poghosyan}. We hope that our contribution may stimulate further research in this direction.

%========================================================================================
\section*{Technical details}
%========================================================================================

The presented simulation results were obtained using a 64-bit version of the four-tap shift-register random number generator introduced by R. M. Ziff~\cite{Ziff} as well as {\tt ran2} from the Numerical Recipes in C~\cite{NRC}. In both cases we have obtained identical results.